\documentclass[8.5pt,twoside]{article}
\usepackage[super,sort&compress,comma]{natbib}
\usepackage{mhchem}
\usepackage{times,mathptmx}
\usepackage{sectsty}
\usepackage{balance}
\usepackage{extarrows}
\usepackage{graphicx} 
\usepackage{lastpage}
\usepackage[format=plain,justification=raggedright,singlelinecheck=false,font=small,labelfont=bf,labelsep=space]{caption}
\usepackage{fancyhdr}
\usepackage{pdfpages}
\usepackage{color}
\newcommand{\ib}[1]{{\color{black}#1}}
\newcommand{\oea}{Opodi \emph{et al.}}
\DeclareGraphicsExtensions{.eps}

\begin{document}

\thispagestyle{plain}
\fancypagestyle{plain}{
\renewcommand{\headrulewidth}{1pt}}
\renewcommand{\thefootnote}{\fnsymbol{footnote}}
\renewcommand\footnoterule{\vspace*{1pt}%
\hrule width 3.4in height 0.4pt \vspace*{5pt}}
\setcounter{secnumdepth}{5}

\makeatletter
\def\subsubsection{\@startsection{subsubsection}{3}{10pt}{-1.25ex plus -1ex minus -.1ex}{0ex plus 0ex}{\normalsize\bf}}
\def\paragraph{\@startsection{paragraph}{4}{10pt}{-1.25ex plus -1ex minus -.1ex}{0ex plus 0ex}{\normalsize\textit}}
\renewcommand\@biblabel[1]{#1}
\renewcommand\@makefntext[1]%
{\noindent\makebox[0pt][r]{\@thefnmark\,}#1}
\makeatother
\renewcommand{\figurename}{\small{Fig.}~}
\sectionfont{\large}
\subsectionfont{\normalsize}

\fancyfoot{}
\fancyfoot[RO]{\footnotesize{\sffamily{1--\pageref{LastPage} ~\textbar  \hspace{2pt}\thepage}}}
\fancyfoot[LE]{\footnotesize{\sffamily{\thepage~\textbar\hspace{3.45cm} 1--\pageref{LastPage}}}}
\fancyhead{}
\renewcommand{\headrulewidth}{1pt}
\renewcommand{\footrulewidth}{1pt}
\setlength{\arrayrulewidth}{1pt}
\setlength{\columnsep}{6.5mm}
\setlength\bibsep{1pt}
\newcommand{\paper}{paper}
\newcommand{\alt}{\raisebox{-0.3ex}{$\stackrel{<}{\sim}$}}
\newcommand{\agt}{\raisebox{-0.3ex}{$\stackrel{>}{\sim}$}}
\newcommand{\figname}{Fig.~}

    \noindent\LARGE{\textbf{Comment on ``A single level tunneling model for molecular junctions: evaluating the simulation methods''
}}
\vspace{0.6cm}

\noindent\large{\textbf{Ioan B\^aldea 
\textit{$^{a \ast}$}
}}\vspace{0.5cm}

\vspace{0.6cm}

\noindent 
\normalsize{Abstract:\\
  The present Comment demonstrates important flaws of the paper Phys. Chem. Chem. Phys. 2022, 24, 11958 by Opodi~\emph{et al.}
  Their crown result (``applicability map'') aims at indicating parameter ranges wherein two approximate methods (called method 2 and 3)
  apply. My calculations reveal that the applicability map is a factual error.
  Deviations of $I_2$ from the exact current $I_1$ do not exceed 3\%
  for model parameters where Opodi~\emph{et al.} claimed that method 2 is inapplicable.
  As for method 3, the parameter range of the applicability map is beyond its scope,
  as stated in papers cited by Opodi~\emph{et al.}~themselves.
}
$ $ \\  

\vspace{0.5cm}

\footnotetext{\textit{$^{a}$~Theoretical Chemistry, Heidelberg University, Im Neuenheimer Feld 229, D-69120 Heidelberg, Germany}}
\footnotetext{$^\ast$~E-mail: ioan.baldea@pci.uni-heidelberg.de
}
Comparing currents $I_1$, $I_2$, and $I_3$ 
through tunneling molecular junctions computed via three single level models (see below),
Opodi \emph{et al.}\cite{Opodi:22} claimed, \emph{e.g.},~that:\\[0.4ex]
\indent (i) The applicability of the method based on $I_3$,\cite{Baldea:2012a} which was previously validated against experiments on 
benchmark molecular junctions (\emph{e.g.}, ref~\citenum{Baldea:2015b,Baldea:2019d,Baldea:2019h})
is ``quite limited''.
\\[0.4ex]
\indent (ii) The ``applicability map'' (Fig.~5 of ref~\citenum{Opodi:22}) should be used in practice
as guidance for the applicability of methods 2 and 3 (\emph{i.e.}, based on $I_2$ and $I_3$) because
(ii$_1$) not only method 3 (ii$_2$) but also method 2 is drastically limited.\\[0.4ex]
\indent (iii) Model parameters for molecular junctions previously extracted from
experimental $I$-$V$-data need revision.

Before demonstrating that these claims are incorrect,
let me briefly summarize
the relevant information available prior to ref~\citenum{Opodi:22}.
Unless otherwise noted
(\emph{e.g.}, the difference between $\Gamma$ and $\tilde{\Gamma}$ expressed by eqn~(\ref{eq-factor-Gamma})),
I use the same notations as ref~\citenum{Opodi:22}.
\begin{equation}
  \label{eq-I1}
  I_1 = 
  \frac{2 e}{h} \left( N \Gamma_{g}^2 \right) \int_{-\infty}^{\infty}
  \frac{f\left(\varepsilon - e V/2\right) - f\left(\varepsilon + e V/2 \right)}{\left(\varepsilon - \varepsilon_0\right)^2 + \Gamma_{a}^2} d\!\varepsilon 
\end{equation}
\begin{equation}
  \label{eq-I2}
  I_2 = \frac{2 e}{h \Gamma_a} \left( N\Gamma_{g}^2\right)
  \left(
  \tan^{-1}\frac{\varepsilon_0 + eV/2}{\Gamma_{a}} -
  \tan^{-1}\frac{\varepsilon_0 - eV/2}{\Gamma_{a}}
  \right)
\end{equation}
\begin{equation}
  \label{eq-I3}
  I_3 = \frac{2 e}{h} \left( N \Gamma_{g}^2\right) \frac{e V}{\varepsilon_0^2 - (e V/2)^2}
  = \underbrace{G_0 \frac{\left( N \Gamma_{g}^2\right)}{\varepsilon_0^2}}_{G_3} \frac{\varepsilon_0^2 V}{\varepsilon_0^2 - (e V/2)^2}
\end{equation}

(a) As a particular case of a formula \cite{Caroli:71a},
eqn~(\ref{eq-I1}) expresses the exact current in the coherent tunneling regime
through a junction consisting of $N$ molecules (set to $N=1$ unless otherwise specified)
mediated by a single level whose energy offset relative to the electrodes' unbiased
Fermi energy is $\varepsilon_0$, coupled to wide, flat band electrodes (hence Lorentzian transmission).
The effective level coupling to electrodes $\Gamma_{g} \equiv \sqrt{\Gamma_s \Gamma_t}$
is expressed in terms of energy independent quantities $\Gamma_{s,t}$,
representing the level couplings to the two electrodes--- substrate (label $s$) and tip (label $t$)---,
which also contribute to the finite level width $\Gamma_{a} = \left(\Gamma_s + \Gamma_t\right)/2$.

In the symmetric case assumed following Opodi \emph{et al.}
\begin{equation}
  \label{eq-sym}
\tilde{\Gamma} = \Gamma_s = \Gamma_t = \Gamma_g = \Gamma_a
\end{equation}
and $\varepsilon_0$ is independent of bias.

In this Comment, I use the symbol $\tilde{\Gamma}$--- a quantity denoted by $\Gamma$
in ref~\citenum{Baldea:2012a} and in all studies on
junctions fabricated with the conducting probe atomic force microscopy (CP-AFM) platform
cited in ref~\citenum{Opodi:22}---
in order to distinguish it from the quantity denoted by $\Gamma$
by Opodi \emph{et al.}\cite{Opodi:22}
Comparison of the present eqn~(\ref{eq-I2}) and (\ref{eq-I3})--- in ref~\citenum{Baldea:2012a} these are
eqn~(3) and (4), respectively--- with eqn~(2) and (3) of ref~\citenum{Opodi:22} makes it clear why:
$\tilde{\Gamma}$ is one half of the quantity denoted by $\Gamma = \Gamma_L + \Gamma_R = 2 \Gamma_L = 2 \Gamma_R $
by Opodi \emph{et al.}\cite{Opodi:22}
\begin{equation}
  \label{eq-factor-Gamma}
\tilde{\Gamma} = \Gamma / 2  
\end{equation}

(b) Eqn~(\ref{eq-I2}) follows as an exact result from eqn~(\ref{eq-I1}) in the zero temperature limit ($T\to 0$),
when the Fermi distribution 
$f(\varepsilon) \equiv 1/\left[1 + \exp\left(\varepsilon / k_B T\right)\right]$
reduces to the step function.

This low temperature limit
(expressed by eqn~(\ref{eq-e0-T}) and (\ref{eq-Gamma-T}) below)
assumes a negligible variation of
the transmission function (which is controlled by $\varepsilon_0$ and $\tilde{\Gamma}$) 
within energy ranges of widths $\sim k_B T$ around the electrodes' Fermi level
wherein electron states switch between full ($f \approx 1$) and
empty ($f\approx 0$) occupancies.
Away from resonance, $\tilde{\Gamma}$ (usually a small value,
$\tilde{\Gamma} \ll \left\vert\varepsilon_0 \right\vert$,
see eqn~(\ref{eq-Gamma-vs-e0}))
plays a negligible role and 
\begin{subequations}
  \label{eq-low-T}
\begin{equation}
  \label{eq-e0-T}
\left\vert\varepsilon_0 \pm eV/2\right\vert \gg k_B T
\end{equation}
is sufficient for the low temperature limit to apply.\cite{Lambert:11}
However, closer to resonance the aforementioned weakly energy-dependent transmission also implies
a sufficiently large $\tilde{\Gamma}$. This implies a relationship between the transmission width ($\sim\tilde{\Gamma}$)
and the width ($\sim k_B T$) of the range ($\sim \left(\varepsilon \pm e V/2 - k_B T, \varepsilon \pm e V/2 + k_B T\right)$, \emph{cf.}~\ref{eq-I1}))
wherein the electrode Fermi functions rapidly vary. In practice, very close to resonance,
loosely speaking, this means \cite{Baldea:2017d,Baldea:2022c,Baldea:2022j}
\begin{equation}
  \label{eq-Gamma-T}
   \tilde{\Gamma} \sim  k_B T
\end{equation}
\end{subequations}

(c) Eqn~(\ref{eq-I3}) was \emph{analytically} derived \cite{Baldea:2012a}
from eqn~(\ref{eq-I2}) for sufficiently large arguments of the inverse trigonometric functions
\begin{equation}
  \label{eq-arctan}
  \tan^{-1}(x) \simeq 
  \frac{\pi}{2} - \frac{1}{x} \mbox{ (holds within 1\% for } x > x_0 = 2.929)
\end{equation}
The bias range wherein eqn~(\ref{eq-I3}) holds within the above accuracy can be expressed as follows
\begin{equation}
  \label{eq-tmp}
  e \vert V \vert < 2 \left(\left\vert \varepsilon_0\right\vert - x_0 \tilde{\Gamma}\right)
  \simeq 2 \left\vert \varepsilon_0\right\vert \left(1 - x_0 \sqrt{g}\right)
\end{equation}
where
\begin{equation}
  \label{eq-g}
  g = \frac{1}{N}\frac{G}{G_0} = \frac{\tilde{\Gamma}^2}{\varepsilon_0^2 + \tilde{\Gamma}^2}
  = \underbrace{\frac{\tilde{\Gamma}^2}{\varepsilon_0^2}}_{g_3}
  \left[1 + \mathcal{O}\left(\frac{\tilde{\Gamma}^2}{\varepsilon_0^2}\right)\right]
\end{equation}
is the zero-temperature low bias conductance \emph{per molecule} ($G/N$)
in units of the universal conductance quantum $G_0 = 2 e^2/h = 77.48\,\mu$S.
Strict on-resonance ($\varepsilon_0 \equiv 0$) single-channel transport is characterized by $g \equiv 1$.
In the vast majority of molecular junctions fabricated so far, tunneling transport
is off-resonant ($g \ll 1$), and $ g < g_{max} = 0.01$ safely holds
in all experimental situations of which I am aware, including all CP-AFM junctions considered in ref~\citenum{Opodi:22}. 
Imposing
\begin{equation}
  \label{eq-g-max}
  g < 0.01
\end{equation}
in eqn~(\ref{eq-g}) yields
\begin{equation}
  \label{eq-Gamma-vs-e0}
  \tilde{\Gamma} < 0.1005 \left \vert \varepsilon_0\right\vert \simeq \left\vert \varepsilon_0\right\vert / 10
\end{equation}
and \emph{via} eqn~(\ref{eq-tmp})\cite{Baldea:2015b,Baldea:2017g}
\begin{subequations}
\label{eq-low-V}
  \begin{eqnarray}
  \label{eq-low-bias}
  \vert e V \vert & < & 1.4\,\left\vert \varepsilon_0\right\vert \mbox{ or, equivalently}\\
  \label{eq-low-bias-vt}
  \vert e V \vert & < & 1.25\, e V_t \mbox{ (\emph{cf.}~eqn~(\ref{eq-vt0}))}
\end{eqnarray}
\end{subequations}
Along with the low-$T$ limit assumed by eqn~(\ref{eq-I2}),
eqn~(\ref{eq-Gamma-vs-e0}) and (\ref{eq-low-bias}) are necessary conditions for eqn~(\ref{eq-I3}) to apply.

Aiming at aiding experimentalists interested in $I$-$V$ data processing, who do not know
$\varepsilon_0$ \emph{a priori}, in ref~\citenum{Baldea:2012a} I rephrased eqn~(\ref{eq-arctan}) by saying that
eqn~(\ref{eq-I3}) holds for biases not much larger than the transition voltage $V_t$ (eqn~(\ref{eq-low-bias-vt})).
$V_t$ is a quantity that can be directly extracted from experiment without any theoretical
assumption from the maximum of $V^2/\vert I\vert$ plotted \emph{vs} $V$.\cite{Beebe:06,Baldea:2015b}
The fact that eqn~(\ref{eq-I3}) should be applied \emph{only}
for biases compatible with eqn~(\ref{eq-low-bias}) has been steadily emphasized (\emph{e.g.},
ref~\citenum{Baldea:2012b} and discussion related
to Fig.~2 and 3, and eqn~4 of ref~\citenum{Baldea:2017g}).

(d) Eqn~(\ref{eq-I3}) is particularly useful because it allows expression of the transition voltage $V_t$
in terms of the level offset \cite{Baldea:2012a}
\begin{equation}
  \label{eq-vt0}
  e V_t = 
  2 \left\vert \varepsilon_0 \right\vert/\sqrt{3}
\end{equation}
which can thus be easily estimated. 
$\varepsilon_0$ is a key quantity in discussing the structure-function relationship in molecular electronics.
Thermal corrections to eqn~(\ref{eq-vt0}), which are significant for small offsets ($\left\vert \varepsilon_0 \right\vert \alt 0.4$\,eV)
even at the room temperature ($k_B T = 25$\,meV) assumed by Opodi \emph{et al.}~were
also quantitatively analyzed (\emph{e.g.}, Fig.~4 in ref~\citenum{Baldea:2017g}).

\emph{To sum up, method 2 applies to situations compatible with
eqn~(\ref{eq-low-T}), and method 3 applies in situations compatible with eqn~(\ref{eq-low-T}) and (\ref{eq-low-V}).
This is a conclusion of a general theoretical analysis that needs no additional confirmation from numerical calculations like those
of ref~\citenum{Opodi:22}.}

Switching to the above claims, 
the following should be said:

\underline{To claim (i)}:

Fig.~2 of ref~\citenum{Opodi:22} shows (along with $\vert I_{1,2} \vert$ also) currents $\vert I_{3} \vert$ computed 
for biases $-1.5\,\mbox{V} < V < +1.5\,\mbox{V}$ at couplings $\Gamma = 2 \tilde{\Gamma} = \{1; 5; 10; 100\}$\,meV
and offsets $\varepsilon_0 = \{0.1; 0.5; 1\}$\,eV.
The lower cusps visible there depict currents
vanishing ($I \to 0, \log\vert I\vert \to -\infty$) at $V = 0$.
The upper symmetric cusps ($\log\vert I\vert \to +\infty$ at $V \to \pm 2 \varepsilon_0 $)
depicted by the blue lines were obtained by \emph{mathematical} application of eqn~(\ref{eq-I3})
beyond the \emph{physically meaningful} bias range of eqn~(\ref{eq-low-bias}), for which eqn~(\ref{eq-I3}) 
was theoretically deduced.

\begin{figure*}[htb]
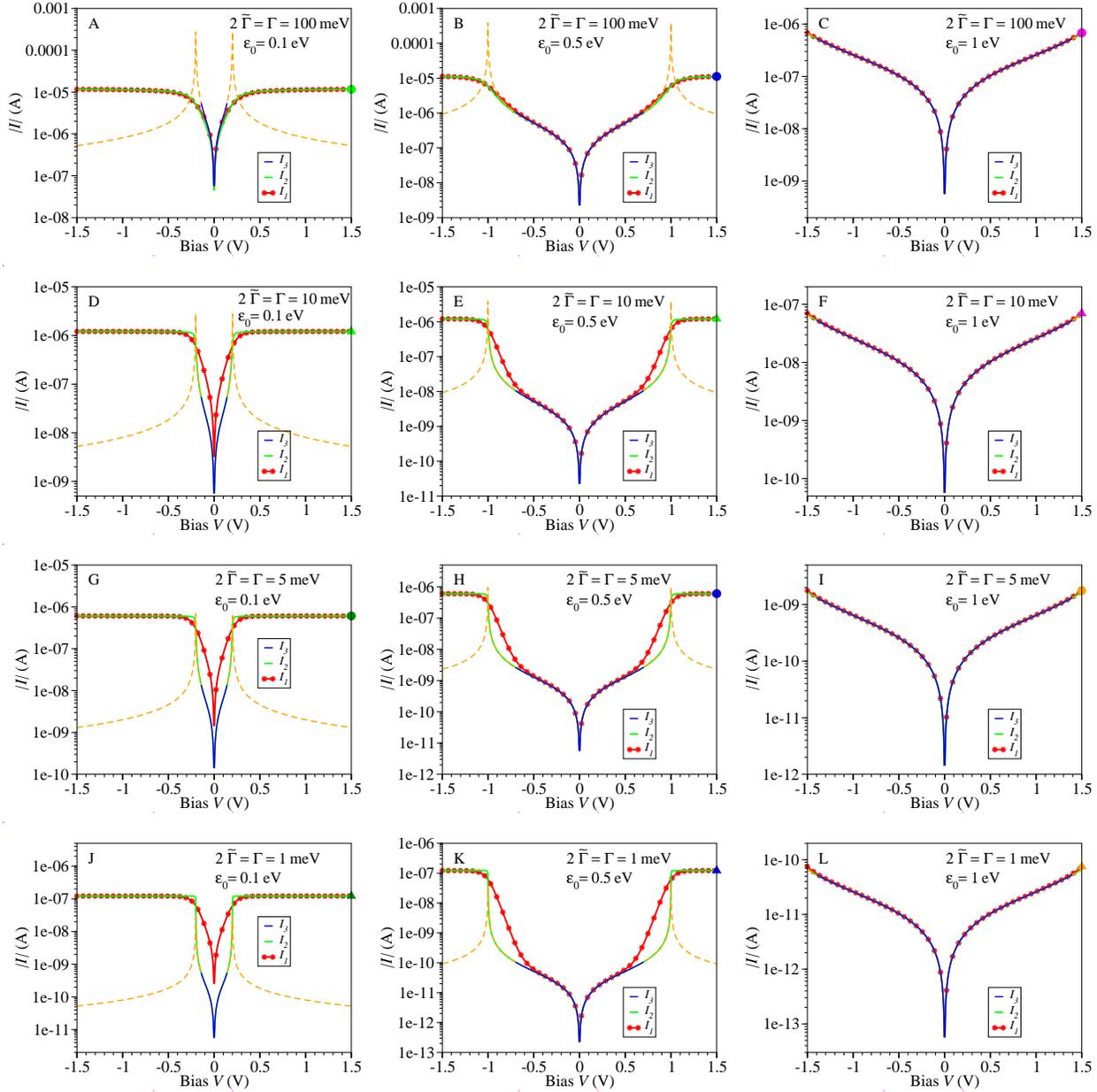

  \centerline{
    \includegraphics[width=0.3\textwidth,angle=0]{Fig2a}
    \includegraphics[width=0.3\textwidth,angle=0]{Fig2b}
    \includegraphics[width=0.3\textwidth,angle=0]{Fig2c}
  }
  \centerline{
    \includegraphics[width=0.3\textwidth,angle=0]{Fig2d}
    \includegraphics[width=0.3\textwidth,angle=0]{Fig2e}
    \includegraphics[width=0.3\textwidth,angle=0]{Fig2f}
  }
  \centerline{
    \includegraphics[width=0.3\textwidth,angle=0]{Fig2g}
    \includegraphics[width=0.3\textwidth,angle=0]{Fig2h}
    \includegraphics[width=0.3\textwidth,angle=0]{Fig2i}
  }
  \centerline{
    \includegraphics[width=0.3\textwidth,angle=0]{Fig2j}
    \includegraphics[width=0.3\textwidth,angle=0]{Fig2k}
    \includegraphics[width=0.3\textwidth,angle=0]{Fig2l}
  }   
  \caption{Currents $I_1$, $I_2$, and $I_3$ computed using the parameters of Fig.~2 of Opodi \emph{et al.}.
    Redrawing their figure emphasizes the difference between the current $I_3$ in the bias range
    for which eqn~(\ref{eq-I3}) was theoretically deduced \cite{Baldea:2012a} (blue curves) and $I_3$ computed
    outside of bias range (orange dashed curves), wherein eqn~(\ref{eq-I3}) is merely a mathematical formula without any physical sense.
    Notice that, throughout, the green ($I_2$) and blue ($I_3$) curves excellently agree with the exact red curves ($I_1$)
    precisely in the parameter ranges predicted by theory, 
    \emph{i.e.}~eqn~(\ref{eq-low-T}) and eqn~(\ref{eq-low-bias}).
    The tick symbols at $V=1.5$\,V depicted in all panels emphasize that method 2 is very accurate,
    invalidating thereby the ``applicability map'' shown by Opodi \emph{et al.} in their Fig.~5 (also reproduced in the present Fig.~\ref{fig:errors}a).}
  \label{fig:fig2}
\end{figure*}

To make clear this point, I corrected in the present Fig.~\ref{fig:fig2}
Fig.~2 of ref~\citenum{Opodi:22}. The blue lines in the present Fig.~\ref{fig:fig2} depict
$I_3$ in the bias range $e\vert V\vert < 1.4\,\left\vert\varepsilon_0\right\vert $ for which eqn~(\ref{eq-I3})
was theoretically deduced and for which it makes physical sense.
The dashed orange lines are merely mathematical curves for $I_3$ computed using
eqn~(\ref{eq-I3}) at 
$e\vert V\vert > 1.4\,\left\vert\varepsilon_0\right\vert $, where they have no physical meaning.
``By definition'', these orange curves are beyond the scope of method 3.

The presentation adopted in Fig.~2 by Opodi \emph{et al.}~also masks 
{the fallacy}
of applying eqn~(\ref{eq-I3}) at biases $\vert e V \vert > 2 \left\vert \varepsilon_0\right\vert$.
There, the denominator in the RHS becomes negative and
bias and current have opposite directions (\emph{i.e.}, $I>0$ for $V<0$ and $I<0$ for $V>0$). 
Visible at $\vert e V  \vert = 2\left\vert \varepsilon_0 \right\vert$ are the nonphysical cusps of the blue ($I_3$)
curves in Fig.~2 by ref.~\citenum{Opodi:22}, 
same as the cusps of the orange curves of the present Fig.~\ref{fig:fig2}.

With this correction, the present Fig.~\ref{fig:fig2}
reveals what it should. Namely that, as long as the off-resonance condition of eqn~(\ref{eq-Gamma-vs-e0})
is satisfied ---
\emph{i.e.}, excepting for Fig.~\ref{fig:fig2}A---,
in all other panels 
the blue and green curves ($I_3$ and $I_2$, respectively) practically coincide.
Significant differences between the exact red curve ($I_1$) and the approximate green and blue ($I_2$ and $I_3$, respectively)
are only visible in situations violating the low temperature condition (eqn~(\ref{eq-low-T})):
in panels D, E, G, and H violating eqn~(\ref{eq-Gamma-T}),
and at biases incompatible with eqn~(\ref{eq-e0-T}). 

\underline{To claim (ii)}:

Refuting \underline{claim (ii$_1$)} is {straightforward}. 
Based on their Fig.~5, Opodi \emph{et al.}~cannot make {a} statement on method 3: they consider
parameters $\varepsilon_0 < 1$\,eV at the bias $V = 1.5$\,V($>1.4\,\left\vert \varepsilon_0\right\vert $),
{which is} incompatible with eqn~(\ref{eq-low-bias}).
Noteworthily, the condition expressed by eqn~(\ref{eq-low-bias}) defied by ref~\citenum{Opodi:22}~was clearly
stated in references that Opodi \emph{et al.}~have cited.

\begin{figure}[htb]
  \centerline{\includegraphics[width=0.28\textwidth,angle=0]{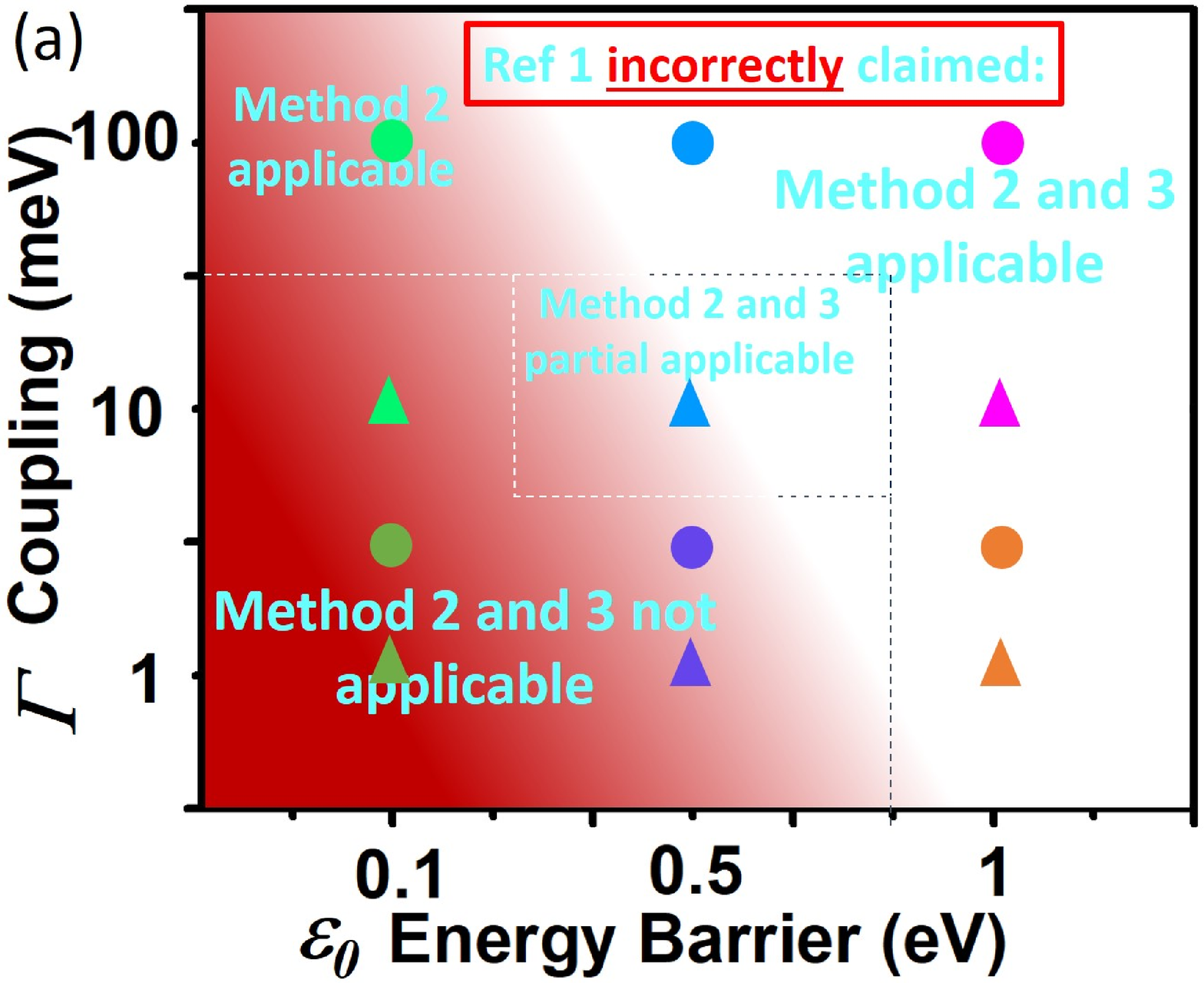}}
  \centerline{\includegraphics[width=0.3\textwidth,angle=0]{fig_error_alongGamma_all}}
  \centerline{\includegraphics[width=0.3\textwidth,angle=0]{fig_error_alongE0_all}}
  \caption{
    (a) Tick symbols depicting excellent agreement between $I_2$ and $I_1$ in all panels of Fig.~\ref{fig:fig2}
      overimposed on the ``applicability map'' adapted after Fig.~5 (courtesy Xi Yu) of ref.~\citenum{Opodi:22}
      contradict the claim of Opodi \emph{et al.} on the inaplicability of method 2.
    (b,c) Deviations of $I_2$ from the exact current $I_1$ 
    reveal that method 2 excellently works in situations where Opodi \emph{et al}~claimed the contrary.
    Importantly, showing parameter values $\left\vert\varepsilon_0\right\vert < 1$\,eV at bias $V=1.5$\,V, panel a is beyond the scope of method 3
    (\emph{cf.}~eqn~(\ref{eq-low-bias})).}
\label{fig:errors}
\end{figure}

To reject \underline{claim (ii$_2$)},
I show in Fig.~\ref{fig:errors}b, c, and d deviations of the current $I_2$ from the exact value $I_1$
in snapshots taken 
horizontally (\emph{i.e.}, constant $\Gamma$) and vertically (\emph{i.e.}, constant $\varepsilon_0$) 
across the ``applicability map'' (\emph{cf.}~Fig.~\ref{fig:errors}a).
As visible in Fig.~\ref{fig:errors}b and c,
in all regions where Opodi \emph{et~al.} claimed that method 2 is invalid
the contrary is true; the largest relative deviation $ I_2 / I_1 - 1$ does not exceed 3\%. 

What the physical quantity is underlying the color code depicted in their Fig.~5 (reproduced here in Fig.~\ref{fig:errors}a)
is not {explained by Opodi \emph{et~al.}}
Anyway, the conclusion of Opodi \emph{et~al.}~summarized in their Fig.~5 contradicts their results shown in their Fig.~2;
all panels of that figure reveal excellent agreement between the green ($I_2$) and red (exact $I_1$)
curves at $V = 1.5$\,V.
For the reader's convenience, the thick symbols at $V = 1.5$\,V in the present Fig.~\ref{fig:fig2} overlapped on  the
``applicability map''of ref~\citenum{Opodi:22} emphasize this aspect.
Inspection of these symbols (indicating that method 2 is excellent) overimposed on Fig.~\ref{fig:errors}a
reveals that they (also) lie in regions where
Opodi \emph{et~al.}~claimed that method 2 fails.
Once more, their ``applicability map'' is factually incorrect. 

\underline{To claim (iii)}:
In their Fig.~3, 4A, 4B, S1 to S4, and S8 as well as in Table~2
Opodi \emph{et~al.}~made unsuitable comparisons: the values for the CP-AFM junctions
taken from their ref~38, 39, 44, and 57
(present ref~\citenum{Baldea:2018a,Baldea:2019d,Baldea:2019h,Frisbie:21a})
are values of $\tilde{\Gamma} = \Gamma/2$, while those estimated by themselves are values of $\Gamma = 2 \tilde{\Gamma}$.
Confusing $\tilde{\Gamma}$ and $\Gamma$, no wonder that they needed re-fitting of the original $I$-$V$ data.
If they had correctly re-fitted the CP-AFM data using $I_3$, with all the values of 
$N$ given in the original works (namely, their ref~38, 39, 44, and 57,
the present ref~\citenum{Baldea:2018a,Baldea:2019d,Baldea:2019h,Frisbie:21a}), up to minor
inaccuracies inherently arising from digitizing the experimental $I$-$V$-curves, they would have 
reconfirmed the values of $\varepsilon_0$ reported in the original publications, and
would have obtained values of $\Gamma = 2\tilde{\Gamma}$
two times larger than those originally reported for $\tilde{\Gamma}$ (\emph{cf.}~eqn~(\ref{eq-factor-Gamma})).

{I still have to emphasize} a difference of paramount importance between $I$-$V$ data fitting based on 
eqn~(\ref{eq-I1}) and (\ref{eq-I2}) on one side, and eqn~(\ref{eq-I3}) on the other side.
Eqn~(\ref{eq-I1}) and (\ref{eq-I2}) have three independent fitting parameters
$\left(\varepsilon_0, N \Gamma_g^2, \Gamma_a\right) \to
\left(\left\vert\varepsilon_0\right\vert, N {\Gamma}^2, {\Gamma}\right)$
while eqn~(\ref{eq-I3}) has only two independent fitting parameters
$\left(\left\vert\varepsilon_0\right\vert, N \Gamma_g^2\right) \to \left(\left\vert\varepsilon_0\right\vert, N {\Gamma}^2\right)$.

All the narrative on the $N$-$\Gamma$-entanglement 
and wording on ``twin sisters'' used in Sec.~3.4 of the original article clearly reveal
that ref.~\citenum{Opodi:22} overlooked that, when using $I_3$, $N$ and $\Gamma$
are two parameters whose values are \emph{impossible} to separate; they
build a unique fitting parameter $N {\Gamma}^2 \equiv 4 N \tilde{\Gamma}^2$.
Data fitting using $I_3$ and three fitting parameters $\left(\varepsilon_0, \Gamma, N\right)$
has an infinity number of solutions for $\Gamma$ and $N$ but they all yield a unique value of $N {\Gamma}^2$.

Were method 3 ``quite limited'' and the deviations of $I_3$ from $I_1$ or $I_2$ significant,
Opodi \emph{et~al.}~would have been able to determine three best fit parameters
$\left(\left\vert \varepsilon_0\right\vert, \Gamma, N\right)$; at least for the ``most problematic'' junctions
where they claimed important departures of $I_3$-based estimates from those based on $I_1$ and $I_2$.
If this is indeed the case, the value of $N$ can be determined from data fitting \cite{Baldea:2022j}.
Their MATLAB code \ib{(additionally relevant details in the ESI)} clearly reveals how they \ib{arrived at}
showing such differences for
real junctions considered. In that code, they keep $N$ fixed and adjust $ \varepsilon_0 $ and  $ \Gamma$.
As long it is reasonably realistic, an arbitrarily chosen value of $N$ has no impact on
directly measurable properties. It changes the value of $\Gamma$ but neither $N \Gamma^2 \propto G_3 \approx G$
nor the level offset  $ \varepsilon_0 $ changes, because method 3 performs well in almost all real cases.

However, defying available values of $N$ for the CP-AFM junctions to which they referred,
Opodi \emph{et al.}~spoke of values up to $N\sim 10^5$. Employing such artificially large $N$'s
nonphysically reduces $\Gamma$ ($N \Gamma^2 \approx \mbox{constant}$, $\Gamma \propto 1/\sqrt{N}$)
down to values incompatible with eqn~(\ref{eq-Gamma-T}),
arriving thereby at the idea that eqn~(\ref{eq-I3}) no longer applies.

In spot checks, I also interrogated curves shown by  Opodi \emph{et al.}~for single-molecule mechanically controllable
break junctions. I arrived so at the junction of 4,4'-bisnitrotolane (BNT),\cite{Zotti:10} the real junction for
which they claimed the most severe failure of method 3.
If Fig.~S7D to S7F and 4D of ref.~\citenum{Opodi:22} were correct, both $\varepsilon_0 $ and $\Gamma$ based on $I_3$ would be in error by a factor of two.

\begin{figure}[htb]
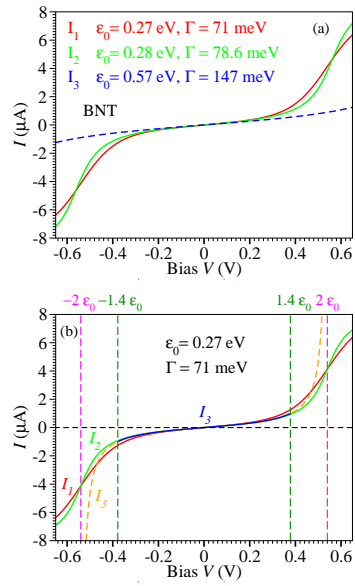

  \centerline{\includegraphics[width=0.28\textwidth,angle=0]{FigS7_iv_BNT}}
    \centerline{\includegraphics[width=0.28\textwidth,angle=0]{FigS7_iv_BNT_ext}}
  \caption{(a)
    $I$-$V$ curves for single-molecule junctions \cite{Zotti:10} obtained using $N=1$ and
    parameters taken from the figures of ref.~\citenum{Opodi:22}
    indicated in the legends.
    To convince himself or herself that the present curves for $I_2$ and $I_3$ are correct and
    different from those of Fig.~S7E and S7E of ref.~\citenum{Opodi:22},
  the reader can easily generate the present green and blue curves by using the GNUPLOT script of the ESI$^\dagger$.
  (b) The curves for $I_1$, $I_2$, and $I_3$ computed with the parameter values (indicated in the inset)
  taken from Fig.~S7D of ref.~\citenum{Opodi:22} do not support the failure of method 3 claimed by {\oea}
}
  \label{fig:mcbj}
\end{figure}

To reject this claim, in Fig.~\ref{fig:mcbj} I show curves for $I_1$, $I_2$, and $I_3$ computed with the values of $ \varepsilon_0 $ and $\Gamma$
indicated by Opodi \emph{et al.} in their Fig.~S7D, E, and F, respectively.
They should coincide with the black curves of Fig.~S7D, E, and F, respectively if the latter were correct.
According to Opodi \emph{et al.}, all these 
would represent fitting curves of the \emph{same} experimental curve (red points in Fig.~S7D, E, and F). 

\ib{Provided that MATLAB is available, the reader
can run the code ``generateIVfitIV.m'' included in the ESI$^\dag$ to convince himself or herself}
that the red curve of Fig.~\ref{fig:mcbj}a and not the black curve of Fig.~S7D
represents the exact current $I_1$ computed using eqn~(\ref{eq-I1}).
\ib{Otherwise,} running the GNUPLOT script also put in the ESI$^\dag$
will at least convince \ib{the} reader 
that the green and the blue lines of this figure and not the black curves in Fig.~S7E and F, respectively
do represent the currents $I_2$ and $I_3$ computed using eqn~(\ref{eq-I2}) and (\ref{eq-I3})
for the parameters and the bias range indicated.
\ib{The} reader will realize that the three curves shown in Fig.~\ref{fig:mcbj} cannot represent best fits of the \emph{same}
experimental $I$-$V$-curve (red points in Fig.~S7D, E, and F of ref~\citenum{Opodi:22}).

If Opodi \emph{et~al.}~had calculated $I_2$ and $I_3$ using the parameters indicated in their Fig.~S7D, E, and F
(same as in the present Fig.~\ref{fig:mcbj}a), they would not have obtained the black curves of their Fig.~S7D, E, and F
but the red, green, and blue curves of Fig.~\ref{fig:mcbj}a. The values $\varepsilon_0 = 0.57$\,eV and $\Gamma = 147$\,meV
of Fig.~S7F (so much different from  $\varepsilon_0 = 0.27$\,eV and $\Gamma = 71$\,meV of Fig.~S7D and $\varepsilon_0 = 0.28$\,eV and $\Gamma = 78.6$\,meV
of Fig.~S7E) can by no means be substantiated from these calculations.
All aforementioned values of $\varepsilon_0$ and $\Gamma$ of Fig.~S7D, E, and F
are exactly the same as the values shown in Table~2 and Fig.~4D of ref~\citenum{Opodi:22}, and used as argument against method 3.

As additional support, I also show (Fig.~\ref{fig:mcbj}b) the curves for $I_1$, $I_2$, and $I_3$, all computed with the same
parameters, namely those of Fig.~S7D of ref~\citenum{Opodi:22} ($\varepsilon_0 = 0.27$\,eV and $\Gamma = 71$\,meV).
In accord with the general theoretical considerations presented under (b) and (c),
the differences between $I_2$ and $I_3$ (blue and green curves) are negligible,
while deviations of $I_2$ from $I_1$ are notable only for biases close $\pm 2\varepsilon_0$ 
that invalidate eqn~(\ref{eq-e0-T}).

To sum up, the claim of Opodi \emph{et~al.}~on the failure of method 3 for the specific case considered above is incorrect because
is based on values incompatible with calculations.

As made clear under (c) above, eqn~(\ref{eq-low-bias}) is a condition deduced analytically. 
If it holds, eqn~(\ref{eq-I3}) is within $\sim 1\%$ as good as eqn~(\ref{eq-I2}).
It makes little sense to check numerically
a general condition deduced analytically,
or even worse (as done in ref.~\citenum{Opodi:22}) to claim that it does not apply for biases incompatible with eqn~(\ref{eq-low-bias}).

The interested scholar needs not the ``applicability map''
(Fig.~5 of ref~\citenum{Opodi:22}, \ib{to be corrected elsewhere (I.~B\^aldea, to be submitted)}).
In the whole parameter ranges where Opodi \emph{et al.}~claimed the opposite, method 2 turned out to be extremely accurate
(\emph{cf.}~Fig.~\ref{fig:errors}b and c).
Likewise, ``by definition'' (cf.~eqn~(\ref{eq-low-bias})), method 3
should not be applied at biases above $ e V > 2 \left\vert\varepsilon_0\right\vert $
shown there, which makes the ``applicability map'' irrelevant for method 3.

Theory should clearly indicate the parameter ranges where an analytic formula is valid.
This is a task accomplished in case of eqn~(\ref{eq-I3}).
In publications also cited by Opodi \emph{et~al.} \cite{Baldea:2015b,Baldea:2017g},
particular attention has been drawn on not to apply eqn~(\ref{eq-I3}) at biases
violating eqn~(\ref{eq-low-bias})\cite{Baldea:2015b} and/or for energy offsets
($\left\vert\varepsilon_0\right\vert \alt 0.5$\,eV) where thermal effects ($k_B T \simeq 25\,\mbox{meV} \neq 0$)
matter \cite{Baldea:2017g}.
It is experimentalists' responsibility not to apply it
under conditions that defy the boundaries under which it was theoretically deduced.

Financial support from the German Research Foundation
(DFG Grant No. BA 1799/3-2) in the initial stage of this work and computational support by the
state of Baden-W\"urttemberg through bwHPC and the German Research Foundation through
Grant No.~INST 40/575-1 FUGG
(bwUniCluster 2.0, bwForCluster/MLS\&WISO 2.0/HELIX, and JUSTUS 2.0 cluster) are gratefully acknowledged.
\renewcommand\refname{Notes and references}
\footnotesize{
\providecommand*{\mcitethebibliography}{\thebibliography}
\csname @ifundefined\endcsname{endmcitethebibliography}
{\let\endmcitethebibliography\endthebibliography}{}

}
\newpage
%
\begin{figure}[htb]
\def\figurename{}
\includegraphics[width=0.90\textwidth,angle=0]{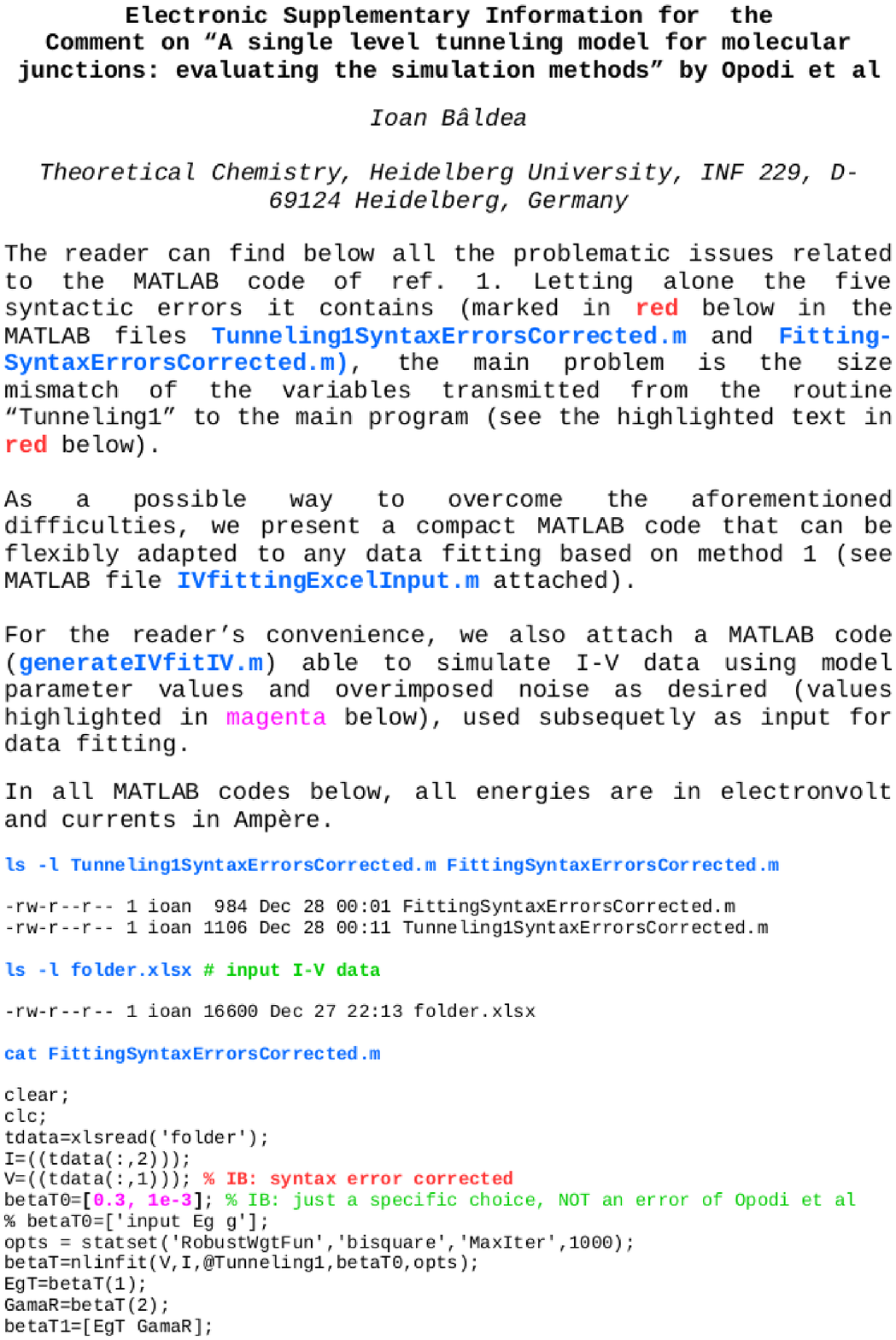}
  \caption{ }
\end{figure}
\newpage
\begin{figure}[htb]
\def\figurename{}
\includegraphics[width=0.99\textwidth,angle=0]{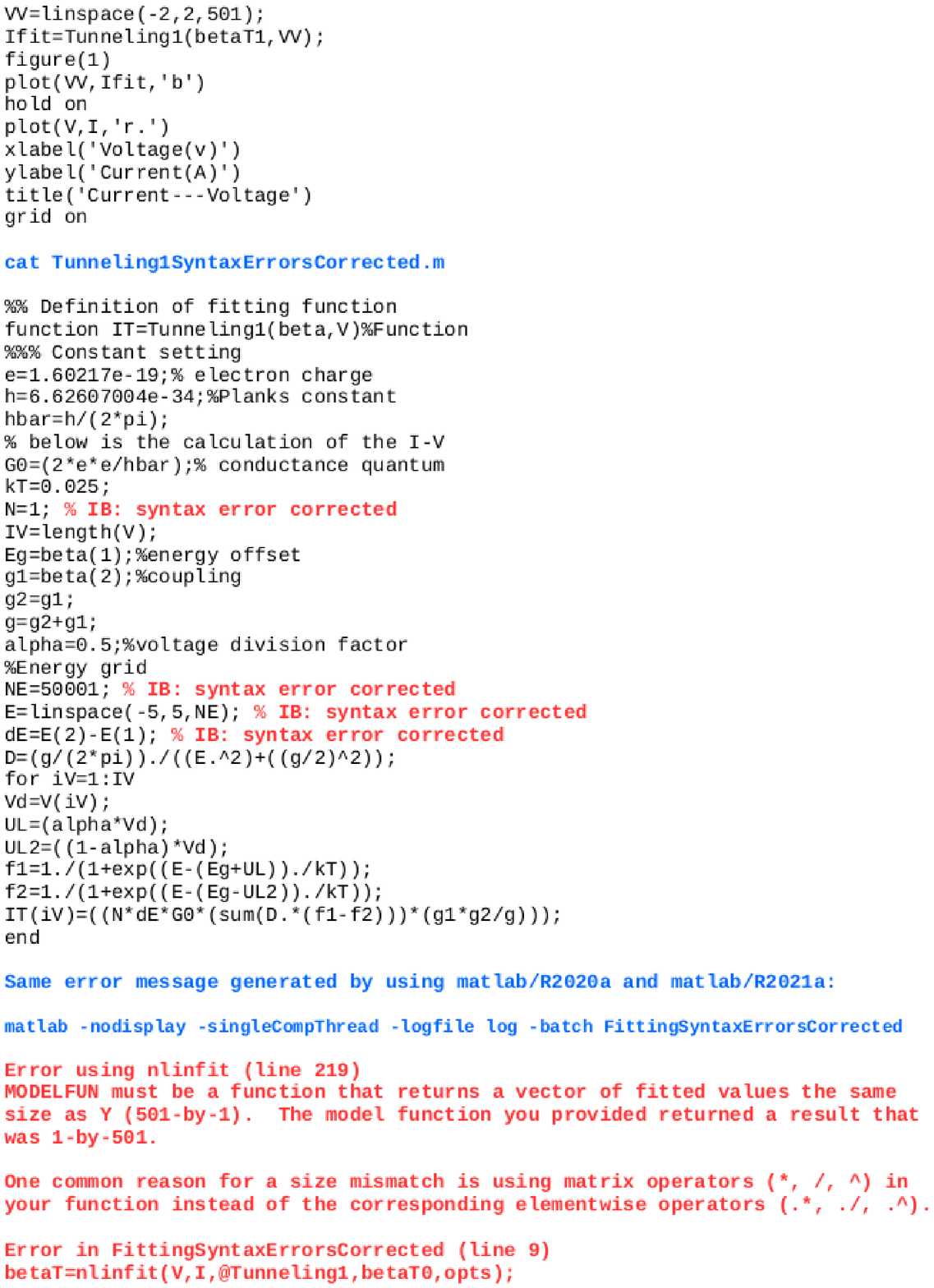}
  \caption{ }
\end{figure}
\begin{figure}[htb]
\def\figurename{}
\includegraphics[width=0.99\textwidth,angle=0]{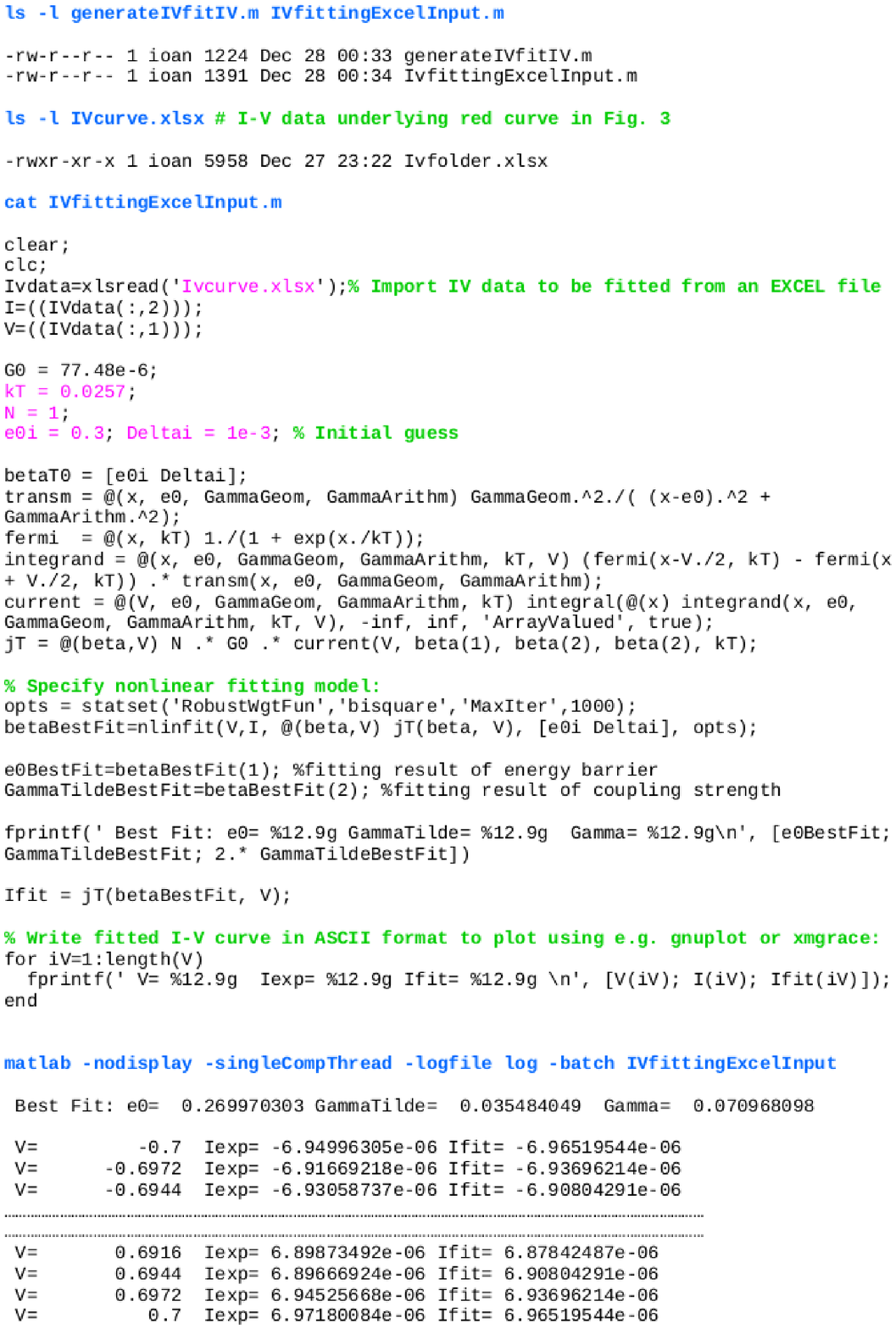}
  \caption{ }
\end{figure}
\begin{figure}[htb]
\def\figurename{}
\includegraphics[width=0.99\textwidth,angle=0]{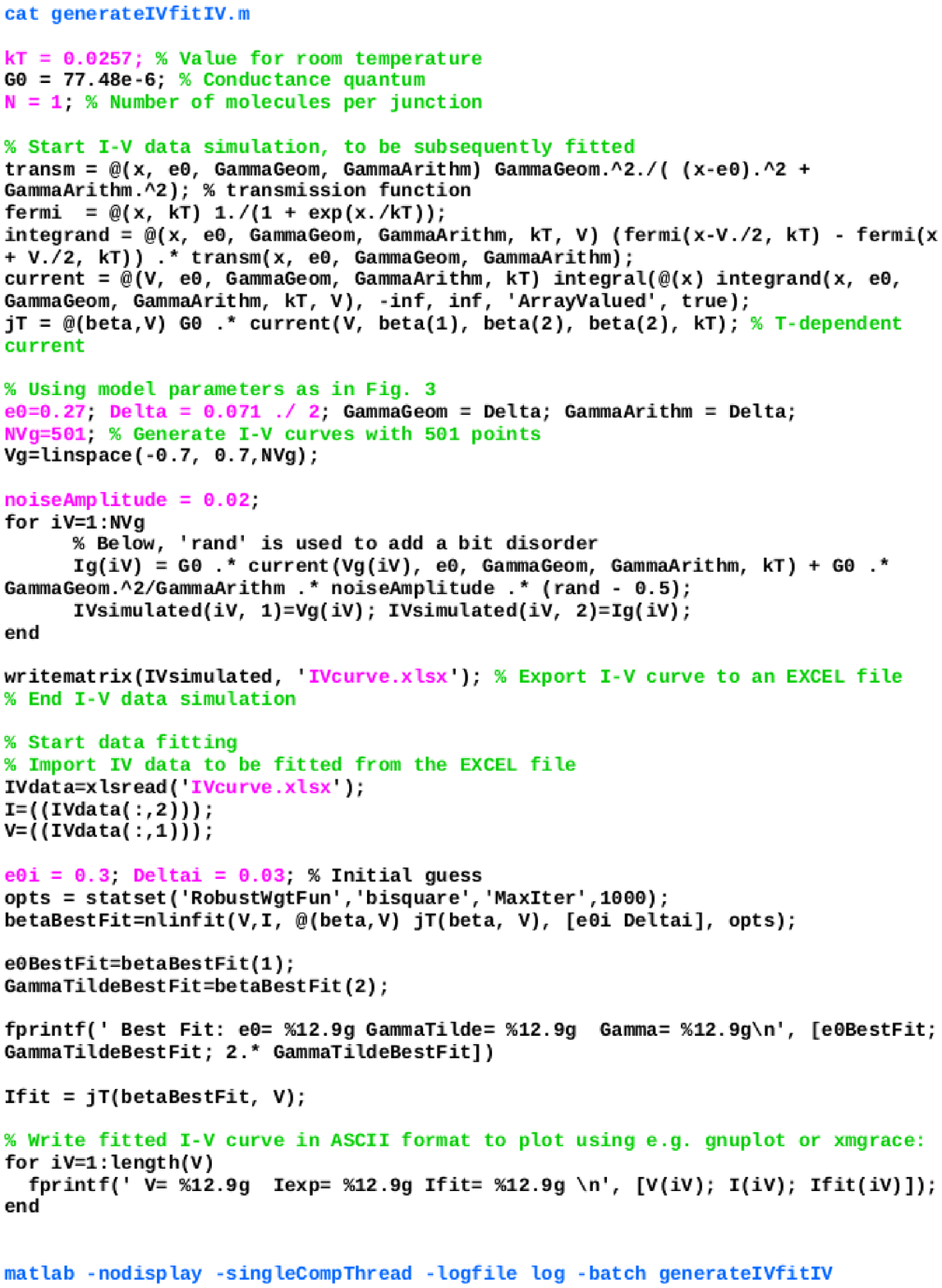}
  \caption{ }
\end{figure}
\begin{figure}[htb]
\def\figurename{}
\includegraphics[width=0.99\textwidth,angle=0]{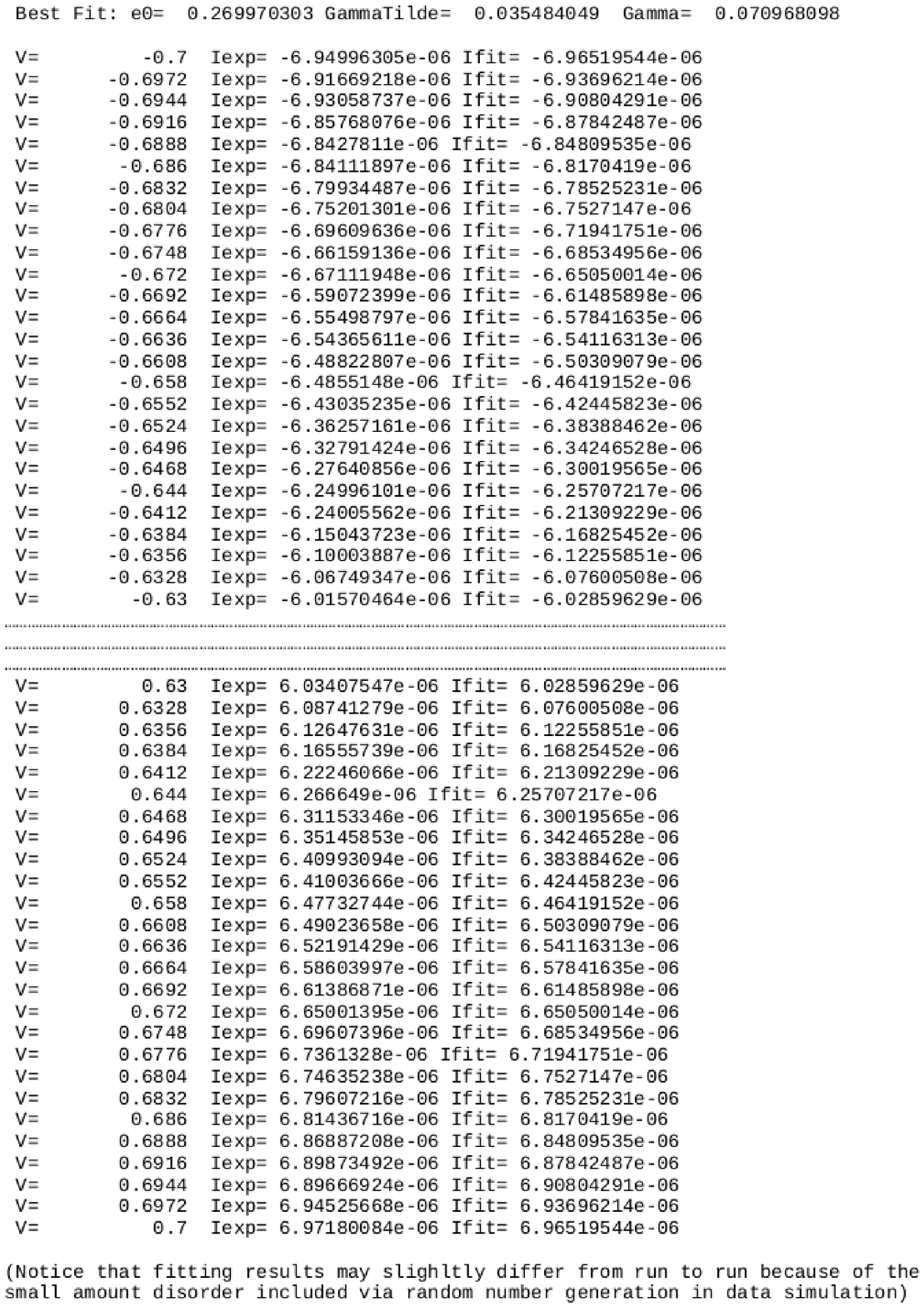}
  \caption{ }
\end{figure}

\end{document}